\documentstyle[twoside]{article}

\catcode`\@=11
\long\def\@makefntext#1{
\protect\noindent \hbox to 3.2pt {\hskip-.9pt
$^{{\eightrm\@thefnmark}}$\hfil}#1\hfill}		

\def\thefootnote{\fnsymbol{footnote}}
\def\@makefnmark{\hbox to 0pt{$^{\@thefnmark}$\hss}}	

\def\ps@myheadings{\let\@mkboth\@gobbletwo
\def\@oddhead{\hbox{}
\rightmark\hfil\eightrm\thepage}
\def\@oddfoot{}\def\@evenhead{\eightrm\thepage\hfil
\leftmark\hbox{}}\def\@evenfoot{}
\def\sectionmark##1{}\def\subsectionmark##1{}}

\oddsidemargin=\evensidemargin
\renewcommand{\thefootnote}{\fnsymbol{footnote}}

\newcounter{sectionc}\newcounter{subsectionc}\newcounter{subsubsectionc}
\renewcommand{\section}[1] {\vspace{12pt}\addtocounter{sectionc}{1}
\setcounter{subsectionc}{0}\setcounter{subsubsectionc}{0}\noindent
	{\tenbf\thesectionc. #1}\par\vspace{5pt}}
\renewcommand{\subsection}[1] {\vspace{12pt}\addtocounter{subsectionc}{1}
	\setcounter{subsubsectionc}{0}\noindent
	{\bf\thesectionc.\thesubsectionc. {\kern1pt \bfit #1}}\par\vspace{5pt}}
\renewcommand{\subsubsection}[1] {\vspace{12pt}\addtocounter{subsubsectionc}{1}
	\noindent{\tenrm\thesectionc.\thesubsectionc.\thesubsubsectionc.
	{\kern1pt \tenit #1}}\par\vspace{5pt}}
\newcommand{\nonumsection}[1] {\vspace{12pt}\noindent{\tenbf #1}
	\par\vspace{5pt}}

\newcounter{appendixc}
\newcounter{subappendixc}[appendixc]
\newcounter{subsubappendixc}[subappendixc]
\renewcommand{\thesubappendixc}{\Alph{appendixc}.\arabic{subappendixc}}
\renewcommand{\thesubsubappendixc}
	{\Alph{appendixc}.\arabic{subappendixc}.\arabic{subsubappendixc}}

\renewcommand{\appendix}[1] {\vspace{12pt}
        \refstepcounter{appendixc}
        \setcounter{figure}{0}
        \setcounter{table}{0}
        \setcounter{lemma}{0}
        \setcounter{theorem}{0}
        \setcounter{corollary}{0}
        \setcounter{definition}{0}
        \setcounter{equation}{0}
        \renewcommand{\thefigure}{\Alph{appendixc}.\arabic{figure}}
        \renewcommand{\thetable}{\Alph{appendixc}.\arabic{table}}
        \renewcommand{\theappendixc}{\Alph{appendixc}}
        \renewcommand{\thelemma}{\Alph{appendixc}.\arabic{lemma}}
        \renewcommand{\thetheorem}{\Alph{appendixc}.\arabic{theorem}}
        \renewcommand{\thedefinition}{\Alph{appendixc}.\arabic{definition}}
        \renewcommand{\thecorollary}{\Alph{appendixc}.\arabic{corollary}}
        \renewcommand{\theequation}{\Alph{appendixc}.\arabic{equation}}
        \noindent{\tenbf Appendix \theappendixc #1}\par\vspace{5pt}}
\newcommand{\subappendix}[1] {\vspace{12pt}
        \refstepcounter{subappendixc}
        \noindent{\bf Appendix \thesubappendixc. {\kern1pt \bfit #1}}
	\par\vspace{5pt}}
\newcommand{\subsubappendix}[1] {\vspace{12pt}
        \refstepcounter{subsubappendixc}
        \noindent{\rm Appendix \thesubsubappendixc. {\kern1pt \tenit #1}}
	\par\vspace{5pt}}

\newcommand{\textlineskip}{\baselineskip=13pt}
\newcommand{\smalllineskip}{\baselineskip=10pt}

\def\eightcirc{
\begin{picture}(0,0)
\put(4.4,1.8){\circle{6.5}}
\end{picture}}
\def\eightcopyright{\eightcirc\kern2.7pt\hbox{\eightrm c}}

\newcommand{\copyrightheading}[1]
	{\vspace*{-2.5cm}\smalllineskip{\flushleft
	{\footnotesize International Journal of Modern Physics B, #1}\\
	{\footnotesize $\eightcopyright$\, World Scientific Publishing
	 Company}\\
	 }}

\newcommand{\publisher}[2]{{\begin{center}\footnotesize\smalllineskip
	Received #1\\
	Revised #2
	\end{center}
	}}

\def\abstracts#1#2#3{{
	\centering{\begin{minipage}{4.5in}\baselineskip=10pt\footnotesize
	\parindent=0pt #1\par
	\parindent=15pt #2\par
	\parindent=15pt #3
	\end{minipage}}\par}}

\renewenvironment{thebibliography}[1]			
	{\frenchspacing
	 \ninerm\baselineskip=11pt
	 \begin{list}{\arabic{enumi}.}
	{\usecounter{enumi}\setlength{\parsep}{0pt}
	 \setlength{\leftmargin 12.7pt}{\rightmargin 0pt} 
	 \setlength{\itemsep}{0pt} \settowidth
	{\labelwidth}{#1.}\sloppy}}{\end{list}}

\newcounter{itemlistc}
\newcounter{romanlistc}
\newcounter{alphlistc}
\newcounter{arabiclistc}

\newcommand{\fcaption}[1]{
        \refstepcounter{figure}
        \setbox\@tempboxa = \hbox{\footnotesize Fig.~\thefigure. #1}
        \ifdim \wd\@tempboxa > 5in
           {\begin{center}
        \parbox{5in}{\footnotesize\smalllineskip Fig.~\thefigure. #1}
            \end{center}}
        \else
             {\begin{center}
             {\footnotesize Fig.~\thefigure. #1}
              \end{center}}
        \fi}

\newcommand{\tcaption}[1]{
        \refstepcounter{table}
        \setbox\@tempboxa = \hbox{\footnotesize Table~\thetable. #1}
        \ifdim \wd\@tempboxa > 5in
           {\begin{center}
        \parbox{5in}{\footnotesize\smalllineskip Table~\thetable. #1}
            \end{center}}
        \else
             {\begin{center}
             {\footnotesize Table~\thetable. #1}
              \end{center}}
        \fi}

\def\@citex[#1]#2{\if@filesw\immediate\write\@auxout
	{\string\citation{#2}}\fi
\def\@citea{}\@cite{\@for\@citeb:=#2\do
	{\@citea\def\@citea{,}\@ifundefined
	{b@\@citeb}{{\bf ?}\@warning
	{Citation `\@citeb' on page \thepage \space undefined}}
	{\csname b@\@citeb\endcsname}}}{#1}}

\newif\if@cghi
\def\cite{\@cghitrue\@ifnextchar [{\@tempswatrue
	\@citex}{\@tempswafalse\@citex[]}}
\def\citelow{\@cghifalse\@ifnextchar [{\@tempswatrue
	\@citex}{\@tempswafalse\@citex[]}}
\def\@cite#1#2{{$\null^{#1}$\if@tempswa\typeout
	{IJCGA warning: optional citation argument
	ignored: `#2'} \fi}}

\def\pmb#1{\setbox0=\hbox{#1}
	\kern-.025em\copy0\kern-\wd0
	\kern.05em\copy0\kern-\wd0
	\kern-.025em\raise.0433em\box0}

\def\fnt#1#2{\footnotetext{\kern-.3em
	{$^{\mbox{\scriptsize #1}}$}{#2}}}

\def\fpage#1{\begingroup
\voffset=.3in
\thispagestyle{empty}\begin{table}[b]\centerline{\footnotesize #1}
	\end{table}\endgroup}

\def\runninghead#1#2{\pagestyle{myheadings}
\markboth{{\protect\footnotesize\it{\quad #1}}\hfill}
{\hfill{\protect\footnotesize\it{#2\quad}}}}
\font\tenrm=cmr10
\font\tenit=cmti10
\font\tenbf=cmbx10
\font\bfit=cmbxti10 at 10pt
\font\ninerm=cmr9
\font\nineit=cmti9
\font\ninebf=cmbx9
\font\eightrm=cmr8

\def\qed{\hbox{${\vcenter{\vbox{			
   \hrule height 0.4pt\hbox{\vrule width 0.4pt height 6pt
   \kern5pt\vrule width 0.4pt}\hrule height 0.4pt}}}$}}

\renewcommand{\thefootnote}{\fnsymbol{footnote}}	

\def\bsc{{\sc a\kern-6.4pt\sc a\kern-6.4pt\sc a}}	
\def\bflatex{\bf L\kern-.30em\raise.3ex\hbox{\bsc}\kern-.14em
T\kern-.1667em\lower.7ex\hbox{E}\kern-.125em X}

\begin{document}

\runninghead{Duval and Horv\'athy
} {Exotic Galilean symmetry and the Hall Effect}

\normalsize\textlineskip
\thispagestyle{empty}
\setcounter{page}{1}

\copyrightheading{}			

\vspace*{0.88truein}

\fpage{1}
\centerline{\bf EXOTIC GALILEAN SYMMETRY AND THE HALL EFFECT
\footnote{Talk given by P. A. Horv\'athy at the Joint APCTP-
Nankai Symposium. Tianjin (China), Oct.2001}}
\vspace*{0.37truein}
\centerline{\footnotesize C. DUVAL}
\vspace*{0.015truein}
\centerline{\footnotesize\it Centre de Physique Th\'eorique, CNRS}
\baselineskip=10pt
\centerline{\footnotesize\it Luminy, Case 907,
F-13 288 MARSEILLE (France)}
\vspace*{10pt}
\centerline{\normalsize and}
\vspace*{10pt}
\centerline{\footnotesize P. A. HORV\'ATHY}
\vspace*{0.015truein}
\centerline{\footnotesize\it Laboratoire de
Math\'ematiques et de Physique Th\'eorique, Universit\'e de Tours}
\baselineskip=10pt
\centerline{\footnotesize\it Parc de Grandmont,
F-37 200 TOURS (France)}
\vspace*{0.225truein}
\publisher{(received date)}{(revised date)}

\vspace*{0.21truein}
\abstracts{The ``Laughlin'' picture of the Fractional Quantum Hall effect
    can be derived using the ``exotic'' model based on the
    two-fold centrally-extended planar Galilei group.
When coupled to a planar magnetic field of critical strength
determined by the extension parameters,
the system becomes singular, and ``Faddeev-Jackiw''
reduction yields the ``Chern-Simons'' mechanics of Dunne, Jackiw, and
Trugenberger. The reduced  system  moves according to the Hall law.}{}{}



\vspace*{1pt}\textlineskip	
\section{Introduction}	
\vspace*{-0.5pt}
\noindent
In his seminal paper \cite{LAUGH}  Laughlin argued that
the Fractional Quantum Hall Effect \cite{QHE}
could be explained as condensation into a collective ground state,
represented by the lowest-Landau-level wave functions
\begin{equation}
    f(z)e^{-\vert z\vert^2/4},
\label{Laughlinwf}
\end{equation}
where the complex $N$-vector $z$ denotes the  positions
of $N$ polarized electrons in the plane; $f(z)$ is analytic.
The fundamental operators are
$\widehat{z}f=zf$, and
$\widehat{\bar{z}}f=2\partial_{z}f$ satisfy
$\big[\widehat{\bar{z}},\widehat{z}\big]=2$.
The quantum Hamiltonian only involves
the potential $V(\bar{z},z)$ suitably quantized with the choice of an
ordering for the non-commuting operators $\widehat{\bar{z}}$ and
$\widehat{z}$.

Our results \cite{DH} presented here say
that the Laughlin picture can actually be obtained from {\it first principles},
namely using the two-fold central extension of the planar Galilei group.
 This latter  has been known for some time \cite{LL} \cite{doubleGal},
 but has long remained a kind of curiosity,
since it had no obvious physical use:
for a free particle of mass $m$, the extra
structure related to the new invariant $k$ leaves the usual
motions unchanged, and only contributes to the conserved quantities
\cite{DH} \cite{doubleGal} \cite{LSZ}.
Let us mention that our ``exotic'' theory
is in fact equivalent to Quantum
Mechanics in the non-commu\-ta\-tive plane \cite{noncommQM},
with non-commutative parameter~$\theta=k/m^2$.

Coupling an ``exotic'' particle
to an electromagnetic field,  the two extension
parameters, $k$ and $m$, combine with the magnetic field, $B$,
into an effective mass, $m^*$, given by (\ref{effmass});
when this latter vanishes,
the consistency of the equations of motion requires
that the particle obey the Hall law.
Interestingly, for $m^*=0$, Hamiltonian reduction \cite{FaJa}
yields the
 ``Chern-Simons mechanics'' considered before by
 Dunne, Jackiw and Trugenberger \cite{DJT}.
The reduced theory admits the infinite symmetry of area-preserving
diffeomorphisms, found before for the edge currents of the Quantum Hall
states \cite{Winfty}.

\textheight=7.8truein
\setcounter{footnote}{0}
\renewcommand{\thefootnote}{\alph{footnote}}

\section{Exotic particle in a gauge field}
\noindent
Let us consider the action
\begin{equation}
 \int{(\vec{p}-\vec{A}\,)\cdot d\vec{x}-h\,dt
+
\frac{\theta}{2}\,\vec{p}\times d\vec{p}
},
\label{matteraction}
\end{equation}
where $(V,\vec{A})$ is an electro-magnetic
potential, the Hamiltonian  being
$
h={\vec{p}{\,}^2}/{2m}+V.
$
The term proportional to the non-commutative parameter $\theta$ is
actually equivalent to the acceleration-dependent Lagrangian
of Lukierski et al. \cite{LSZ}.
 The associated Euler-Lagrange equations read
\begin{equation}
\left\{\!
\begin{array}{rcl}\displaystyle
m^*\dot{x}_{i}
&=&
p_{i}-\displaystyle m\theta\,\varepsilon_{ij}E_{j},
\\[8pt]
\displaystyle
\dot{p}_{i}
&=&
E_{i}+B\,\varepsilon_{ij}\dot{x}_{j},
\end{array}
\right.
\label{eqmotion}
\end{equation}
where we have introduced the \textit{effective mass}
\begin{equation}
m^*=m(1-\theta B).
\label{effmass}
\end{equation}
The velocity and momentum
are different if $\theta\neq0$.
The equations of motions (\ref{eqmotion}) can also be written as
\begin{equation}
	\omega_{\alpha\beta}\dot{\xi}_\beta=\frac{\partial h}{\partial
\xi_\alpha},
\qquad\hbox{where}\qquad
	\big(\omega_{\alpha\beta}\big)=
	\left(\begin{array}{cccc}
	0&\theta&1&0\\[2mm]
	-\theta&0&0&1\\[2mm]
	-1&0&0&B\\[2mm]
	0&-1&-B&0\\[2mm]
	\end{array}\right).
\label{symplecticmatrix}
\end{equation}
Note that the electric and magnetic fields
are otherwise arbitrary solutions of the homogeneous Maxwell equation
$\partial_tB+\varepsilon_{ij}\partial_iE_j=0$, which guarantees that the
two-form
$\omega=\frac{1}{2}\omega_{\alpha\beta}d\xi_{\alpha}\wedge{}d\xi_{\beta}$
is closed, $d\omega=0$.

\goodbreak
When  $m^*\neq0$, the determinant
  $\det\big(\omega_{\alpha\beta}\big)=
\left(1-\theta\, B\right)^2
=\big(m^*/m\big)^2$
is nonzero and the matrix  $(\omega_{\alpha\beta})$ in (\ref{symplecticmatrix})
  can  be inverted.
 Then the equations of motion
(\ref{symplecticmatrix}) (or (\ref{eqmotion}))
take the form $\dot{\xi}_{\alpha}=\big\{\xi_{\alpha}, h\big\}$,
with the standard Hamiltonian, but with the new Poisson
bracket
$\{f,g\}=(\omega^{-1})_{\alpha\beta}\partial_{\alpha}f\partial_{\beta}g$
which reads,
explicitly,
\begin{equation}
\begin{array}{rcl}
\{f,g\}
&=&
\displaystyle\frac{m}{m^*}\left[
\frac{\partial f}{\partial\vec{x}}\cdot\frac{\partial g}{\partial\vec{p}}
-
\frac{\partial g}{\partial\vec{x}}\cdot\frac{\partial f}{\partial\vec{p}}
\right.\\[12pt]
&&
+
\left.
{\theta}\left(
\displaystyle\frac{\partial f}{\partial x_1}\frac{\partial g}{\partial x_2}
-
\frac{\partial g}{\partial x_1}\frac{\partial f}{\partial x_2}\right)
+
B\displaystyle\left(
\frac{\partial f}{\partial p_1}\frac{\partial g}{\partial p_2}
-
\frac{\partial g}{\partial p_1}\frac{\partial f}{\partial p_2}\right)
\right].\hfill
\end{array}
\label{exoticPB}
\end{equation}
\goodbreak

Further insight can be gained when the magnetic field $B$
is a (positive) nonzero constant, which turns out the most interesting case,
and will be henceforth assumed.
(The electric field $E_i=-\partial_iV$ is
still arbitrary). Introducing the new coordinates
\begin{equation}
\left\{\begin{array}{c}
Q_{i}=x_{i}+
\displaystyle\frac{1}{B}\left[1-\sqrt{
\displaystyle\frac{m^*}{m}}\,\right]
\varepsilon_{ij}\,p_{j},\\[8pt]
P_{i}=\sqrt{\displaystyle\frac{m^*}{m}}\,p_i-
\displaystyle\frac{1}{2}B\varepsilon_{ij}\,Q_{j},\hfill
\end{array}\right.
\label{goodcoordinates}
\end{equation}
will allow us to generalize our results in
\cite{DH} from a constant to any electric field.

Firstly, the Cartan one-form \cite{SSD} in the action (\ref{matteraction})
reads simply
$
P_idQ_i-hdt,
$
so that the symplectic form on phase space retains the canonical guise,
$\omega=dP_{i}\wedge dQ_{i}$.
The price to pay is that the
Hamiltonian becomes rather complicated \cite{DH}.

The equations of motion (\ref{eqmotion}) are conveniently presented
in terms of the new variables $\vec{Q}$ and the old momenta $\vec{p}$, as
\begin{equation}
    \left\{\begin{array}{ll}
\dot{Q}_{i}=\varepsilon_{ij}\displaystyle\frac{E_{j}}{B}
+\displaystyle\sqrt{\frac{m}{m^*}}\left(
\frac{p_{i}}{ m}-\varepsilon_{ij}\displaystyle\frac{E_{j}}{B}\right),
\\[4mm]
\dot{p}_{i}=
\varepsilon_{ij}{B}\displaystyle\frac{m}{m^*}\left(
\frac{p_{j}}{ m}-
\varepsilon_{jk}\displaystyle\frac{E_{k}}{B}\right).
\end{array}\right.
\label{Qpeqmot}
\end{equation}
Note that all these expressions diverge when
$m^*$ tends to zero.

When the magnetic field takes the particular value
\begin{equation}
B=B_{c}=\frac{1}{\theta},
\label{critB}
\end{equation}
the effective mass (\ref{effmass}) vanishes, $m^*=0$, so that
$\det(\omega_{\alpha\beta})=0$, and the system becomes singular.
Then the time derivatives $\dot{\xi_\alpha}$ can no longer be
expressed from the
variational equations (\ref{symplecticmatrix}), and we have resort to
``Faddeev-Jackiw'' reduction \cite{FaJa}. In accordance with the Darboux
theorem (see, e.g.,~\cite{SSD}), the Cartan one-form
in (\ref{matteraction})
can be written, up to an exact term, as
\begin{equation}
\vartheta-hdt,
\qquad\hbox{with}\qquad
\vartheta=
(p_{i}-\frac{1}{2}{B_{c}}\,\varepsilon_{ij}\,x_{j})dx_{i}
+\frac{1}{2}\theta\varepsilon_{ij}\,p_{i}dp_{j}
=
P_{i}dQ_{i},
\end{equation}
where the new coordinates read, consistently with
(\ref{goodcoordinates}),
\begin{equation}
Q_{i}=x_{i}+\displaystyle\frac{1}{B_{c}}\varepsilon_{ij}p_{j},
\label{redcancoord}
\end{equation}
while the
$P_{i}=-\frac{1}{2}{}B_{c}\,\varepsilon_{ij}\,Q_{j}$
are in fact the rotated coordinates $Q_{i}$.
Eliminating the original coordinates $\vec{x}$ and $\vec{p}$
using (\ref{redcancoord}),
we see that the Cartan one-form reads
$
P_idQ_i-H(\vec{Q},\vec{p})dt,
$
where
$
H(\vec{Q},\vec{p})=\vec{p}{\,}^2/(2m)+V(\vec{Q},\vec{p}).
$
As the $p_{i}$ appear here with no derivatives,
they can be eliminated using their equation of motion
$
\partial H(\vec{Q},\vec{p})/\partial\vec{p}=0,
$
i. e., the constraint
\begin{equation}
\frac{p_i}{m}-\frac{\varepsilon_{ij}E_j}{B_{c}}=0.
\label{pHall}
\end{equation}
A short calculation shows that the reduced
Hamiltonian is just the original potential,
viewed as a function of the ``twisted'' coordinates $\vec{Q}$, {\it viz.}
\begin{equation}
H=V(\vec{Q}).
\label{redham}
\end{equation}
This rule is referred to as the ``Peierls substitution''
\cite{DJT} \cite{DH}.
Since $\partial^2H/\partial{}p_i\partial{}p_j=\delta_{ij}/m$ is
already non singular,
the reduction stops, and we end up with the reduced Lagrangian
\begin{equation}
L_{\rm red}=\frac{1}{2\theta}\vec{Q}\times\dot{\vec{Q}}-V(\vec{Q}),
\label{redlag}
\end{equation}
supplemented with the Hall constraint (\ref{pHall}).
The $4$-dimensional phase space is hence reduced to $2$ dimensions, with
$Q_1$ and $Q_2$ in (\ref{redcancoord})
as canonical coordinates, and reduced symplectic two-form
$
\omega_{\rm red}=
\frac{1}{2} B_{c}\,\varepsilon_{ij}dQ_{i}\wedge dQ_{j}
$
so that the reduced Poisson bracket is
\begin{equation}
\big\{F, G\big\}_{{\rm red}}=-\frac{1}{B_{c}}
\Big(\frac{\partial F}{\partial Q_{1}}\,\frac{\partial G}{\partial Q_{2}}
-\frac{\partial G}{\partial Q_{1}}\,\frac{\partial F}{\partial Q_{2}}\Big).
\label{redpoisson}
\end{equation}
The twisted coordinates are therefore again
non-commuting,
\begin{equation}
\big\{Q_{1}, Q_{2}\big\}_{{\rm red}}=-\theta=-\frac{1}{B_{c}}.
\label{redcommrel}
\end{equation}
The equations of motion associated with (\ref{redlag}), and also consistent
with
the Hamilton equations $\dot{Q}_i=\big\{Q_i,H\}_{{\rm red}}$, are given by
\begin{equation}
\dot{Q}_{i}=\varepsilon_{ij}\frac{E_{j}}{B_{c}},
\label{QHall}
\end{equation}
in accordance with the Hall law (compare (\ref{Qpeqmot}) with the
divergent terms removed).
\goodbreak

Putting $B_{c}=1/\theta$, the Lagrangian (\ref{redlag}) becomes formally
identical to the one
Dunne et al.~\cite{DJT} derived letting the {\it real} mass
go to zero. Note, however, that while $\vec{Q}$ denotes
real position in Ref.~\cite{DJT}, our $\vec{Q}$ here is  the
``twisted'' expression  (\ref{redcancoord}), with the magnetic field
frozen at the critical value $B_{c}=1/\theta$.

\section{Infinite symmetry}
\noindent
It has been argued \cite{Wen} that the physical process
which yields the Fractional Quantum Hall Effect actually takes
place at the boundary of the droplet of the ``Hall'' liquid:  owing
to incompressibility, the bulk can not support any density waves, but there
are chiral currents at the edge.
These latter fall into irreducible representations of the infinite dimensional
algebra $W_{1+\infty}$ \cite{Winfty}, which is the
quantum deformation of $w_{\infty}$, the algebra of classical observables
which generate the group of
area-preserving diffeomorphisms of the plane.

Our reduced model is readily seen to admit $w_{\infty}$,
the classical counterpart
of $W_{1+\infty}$, as symmetry.
To see this, let us remember that, as argued by Souriau \cite{SSD},
and later by Crnkovic and Witten \cite{CW},
it is convenient to consider the space of solutions of the
equations of motion
(Souriau's ``{\it espace des mouvements}'' [= space of motions]),
denoted by ${\cal M}$. For a classical
mechanical system,
this is an abstract substitute for the classical phase space,
whose points are the motion curves of the system. The classical
dynamics is encoded into the symplectic form $\Omega$ of ${\cal M}$.
It is then obvious that {\it any} function $f(\zeta)$ on ${\cal M}$ is
a constant of the motion. (When expressed using
the positions, time, and momenta, such a function can look rather
complicated).
Any such function $f(\zeta)$ generates a Hamiltonian
vectorfield $Z^\mu$ on ${\cal M}$ through the
relation
\begin{equation}
    -\partial_{\mu}f=\Omega_{\mu\nu}Z^\nu.
\end{equation}

The vector field $Z^\mu$ generates, at least locally, a $1$-parameter
group of diffeomorphisms of ${\cal M}$.
All diffeomorphisms of ${\cal M}$ which leave the symplectic form
$\Omega$ invariant form an infinite dimensional group, namely the
group of
symplectomorphisms  of ${\cal M}$. Any symplectic transformation is a
symmetry of the system : it merely permutes the motions curves.

For the reduced system above, the reduced phase space is two
dimensional. The space of motions is therefore  locally a plane.
(Its global structure plainly depends on the details of the dynamics).
Now, for any orientable two dimensional manifold, the symplectic
form is the area element; it follows that the reduced system admits
the group of area-preserving transformations as symmetry.


\section{Quantization}
\noindent
Let us conclude our general theory by  quantizing the coupled system.
Again, owing to the exotic term, the position representation does not exist.

Introducing the complex coordinates
\begin{equation}
\left\{\begin{array}{c}
z=\displaystyle\frac{\sqrt{B}}{2}\big(Q_1+iQ_2\big)
+\displaystyle\frac{1}{\sqrt{B}}\big(-iP_1+P_2)\hfill
\\[3mm]
w=\displaystyle\frac{\sqrt{B}}{2}\big(Q_1-iQ_2\big)
+\displaystyle\frac{1}{\sqrt{B}}\big(-iP_1-P_2)\hfill
\end{array}\right.
\label{bigcomplexcoord}
\end{equation}
the two-form $dP_i\wedge dQ_i$ on $4$-dimensional unreduced
phase space becomes the
canonical K\"ahler two-form of ${\bf C}^2$, viz
$
\omega=(2i)^{-1}\big(d\bar{z}\wedge dz+d\bar{w}\wedge dw\big).
$
 choosing the antiholomorphic polarization, the ``unreduced''
quantum Hilbert space, consisting of
the ``Bargmann-Fock'' wave functions
\begin{equation}
	\psi(z,\bar{z},w,\bar{w})
	=f(z,w)e^{-\frac{1}{4}(z\bar{z}+w\bar{w})},
\label{bigBF}
\end{equation}
where $f$ is holomorphic in both of its variables.
The fundamental quantum operators,
\begin{equation}
	\left\{\begin{array}{ll}
		\widehat{z}\,f=zf,
		&\widehat{\bar{z}}\,f=2\partial_{z}f,
		\\[8pt]
		\widehat{w}f=wf,
		&\widehat{\!\bar{w}}f=2\partial_{w}f,
	\end{array}\right.
	\label{bigmomop}
\end{equation}
satisfy the commutation relations
$\big[\widehat{\bar{z}},\widehat{z}\big]
=\big[\,\widehat{\!\bar{w}},\widehat{w}\big]
=2$,
and
$
\big[\widehat{z},\widehat{w}\big]
=\big[\widehat{\bar{z}},\widehat{\!\bar{w}}\big]
=0
$.
 We recognize here the familiar creation and annihilation operators, namely
$a^*_z=z$, $a^*_w=w$, and $a_z=\partial_z$, $a_w=\partial_w$.
Using (\ref{goodcoordinates}), the (complex)  momentum
$p=p_{1}\!+\! ip_{2}$
and the kinetic part, $h_0$, of the Hamiltonian become,
respectively,
\begin{equation}
	p=-i\sqrt{\frac{mB}{m^*}}\,\bar{w}
	\qquad\hbox{and}\qquad
	h_{0}=
	\frac{B}{2m^*}\,w\bar{w}.
\label{complexmomentum}
\end{equation}

For $m^*\neq0$ the wave function satisfies the Schr\"odinger
equation
$
i\partial_tf=\widehat{h}f,
$
with
$
\widehat{h}=\widehat{h}_{0}+\widehat{V}.
$
The quadratic kinetic term here is
 \begin{equation}
	 \widehat{h}_{0}=
	 \frac{B}{4m^*}\big(\widehat{w}\,\widehat{\!\bar{w}}+
	 \widehat{\!\bar{w}}\,\widehat{w}\big)=
	 \frac{B}{2m^*}\big(\widehat{w}\,\widehat{\!\bar{w}}+1\big).
\label{kinham}
\end{equation}

The case when the effective mass tends to zero is conveniently studied in this
framework. On the one hand, in the limit $m^*\to0$, one has
\begin{equation}
	z\to\sqrt{B}\,Q,
	\qquad
	w\to 0,
\end{equation}
where $Q=Q_{1}+iQ_{2}$, cf. (\ref{goodcoordinates}); the $4$-dimensional
phase space reduces to the complex plane.
On the other hand, from
(\ref{complexmomentum}) and (\ref{bigmomop}) we deduce that
\begin{equation}
	i\sqrt{\frac{m^*}{mB}}\,\widehat{p}=\widehat{\!\bar{w}}=2\partial_{w}.
\end{equation}
The limit $m^*\to0$ is hence enforced, at the quantum level,
by requiring that the wave functions be independent of the
coordinate $w$, i.e.,
\begin{equation}
\partial_{w}f=0,
\label{LLLcond}
\end{equation}
yielding the reduced wave functions of the form
\begin{equation}
\Psi(z,\bar{z})=f(z)e^{-\frac{1}{4}z\bar{z}},
\label{redwavefunction}
\end{equation}
where $f$ is a holomorphic function of the reduced phase space parametrized by
$z$.
When viewed in the ``big'' Hilbert space (see~(\ref{bigBF})), these wave
functions
belong, by (\ref{kinham}), to the lowest Landau level \cite{QHE}
\cite{GJ} \cite{DH}.

Using the fundamental operators $\widehat{z}$ an $\widehat{\bar{z}}$ given in
(\ref{bigmomop}), we easily see that the (complex) ``physical'' position
$x=x_{1}+ix_{2}$ and its quantum counterpart $\widehat{x}$, namely
\begin{equation}
x=\frac{1}{\sqrt{B_c}}\left(z+\sqrt{\frac{m}{m^*}}\,\bar{w}\right),
	\qquad
\widehat{x}=
\frac{1}{\sqrt{B_c}}\left(z+\sqrt{\frac{m}{m^*}}\,2\partial_{w}\right),
\label{complexposition}
\end{equation}
manifestly diverge when $m^*\to0$.
{\it Positing from the outset} the conditions (\ref{LLLcond})
 the divergence is suppressed, however,
 leaving us with the reduced position operators
\begin{equation}
	\widehat{x}\,f=\widehat{Q}f=\frac{1}{\sqrt{B_c}}\,zf,
\qquad
	\widehat{\bar{x}}\,f=\widehat{\!\bar{Q}}f=
	\frac{2}{\sqrt{B_c}}\,\partial_zf,
\label{redquantumposition}
\end{equation}
whose commutator is
$[\widehat{Q},\widehat{\!\bar{Q}}]={2}/{B_c},$
cf. (\ref{redcommrel}).
In conclusion, we recover the
``Laughlin'' description (\ref{Laughlinwf})
of the ground states of the FQHE in  \cite{QHE}.
Quantization of the reduced Hamiltonian (which is, indeed, the
potential $V(z,\bar{z})$),
can be achieved using, for instance, anti-normal ordering  \cite{DJT}
\cite{GJ}.

\nonumsection{Acknowledgements}
\noindent
We are indebted to Professors J. Gamboa, Z. Horv\'ath, R.~Jackiw, L. Martina
and F. Schaposnik for
discussions. P. H. would like to thank Prof. Mo-lin Ge for
hospitality extended to him at  Nankai Institute of Tianjin (China).

\nonumsection{References}

\end{document}